\documentstyle[12pt]{article}

\title{{\bf Study of the Wheeler Propagator}
\thanks{\it{This work was partially supported by Consejo Nacional de
Investigaciones Cient\'{\i}ficas and Comisi\'{o}n de Investigaciones
Cient\'{\i}ficas de la Pcia. de Buenos Aires; Argentina.}}}
\author{C.G.Bollini${}^1$ and M.C.Rocca${}^{1,2}$ \\
${}^1$Departamento de F\'{\i}sica, Fac. de Ciencias Exactas,\\
Universidad Nacional de La Plata.\\
C.C. 67 (1900) La Plata. Argentina. \\
${}^2$Departamento de Matem\'{a}ticas, Fac. de Ciencias Exactas \\
Universidad Nacional del Centro de la Pcia de Bs. As.\\
Pintos 390, C.P.7000 , Tandil. Argentina.}

\date{August 1, 1997}

\begin{document}

\maketitle

\begin{abstract}
We study the half advanced and half retarded Wheeler Green function
and its relation to Feynman propagators. First for massless equation.
Then, for Klein-Gordon equations with arbitrary mass parameters;
real, imaginary or complex. In all cases the Wheeler propagator
lacks an on-shell free propagation. The Wheeler function has support
inside the light-cone ( whatever the mass ). The associated vacuum
is symmetric with respect to annihilation and creation operators.

We show with some examples that perturbative unitarity holds, whatever
the mass ( real or complex ). Some possible applications are
discussed.


\end{abstract}

\newpage

\section{Introduction}

More than half a century ago, J. A. Wheeler and R. P. Feynman plublished
a work \cite{tp1} in which they represented electromagnetic
interactions by means of a half advanced and half retarded potential.
The charged medium was supossed to be a perfect absorber, so that no
radiation could possibly scape the system.

We are going to call this kind of potential a ``Wheeler function''
( or propagator ), although it had been used before by P. A. M. Dirac
\cite{tp2} when trying to avoid some run-away solutions. Later, in
1949, J. A. Wheeler and R. P. Feynman showed that, in spite of the
fact that it contains an advanced part, the results do no contradict
causality \cite{tp3}.

Of course, the success of QED and renormalization theory made soon
unnecesary or not advisable, to follow that line of research
( at least for electromagnetism ).

One of the distinctive characteristics of the Green function used
in references \cite{tp1,tp2,tp3} is its lack of asymptotic free waves.
This is the reason behind the choice of a ``perfect absorber'' for the
medium through which the field propagates. As the quantization of
free waves is associated to free particles, the above mentioned
feature of Wheeler functions imply that no free quantum of the field
can ever be observed. Nevertheless, we are now habituated to the
existence of confined particles. They do not manifiest themselves
as free entities.

We can give some examples ( outside QCD ) where such a behaviour
can be present.

A Lorentz-invariant higher order equation can be decompossed into
Klein-Gordon factors, but the corresponding mass parameters need not
be real. For instance, the equation:

\begin{equation}
\left( {\Box}^2 + m^4 \right) \varphi = \left( \Box + im^2 \right)
\left( \Box - im^2 \right) \varphi = 0
\end{equation}

gives rise to a pair of constituent fields \cite{tp4} obeying:

\begin{equation}
\left( \Box \pm im^2 \right) {\varphi}_{\pm} = 0
\end{equation}

Any solution of eq.(2) blows-up asymptotically. We can say that the
corresponding fields should be forbidden to appear asymptotically
as free waves. Therefore, they should have a Wheeler function as
propagator \cite{tp5}.

Equations similar to (1), or more general:

\begin{equation}
\left( {\Box}^n \pm m^{2n} \right) \varphi = 0
\end{equation}

appear in a natural way in supersymmetric models for higher dimensional
spaces \cite{tp6}.

Another example is provided by fields obeying Klein-Gordon equations
with the wrong sign of the mass term. A careful analysis shows that
the propagator should be a Wheeler function \cite{tp7,tp8}.
Accordingly no tachyon can ever be observed as a free particle.
They can only exist as ``mediators'' of interactions.

To define the propagators in a proper way, we have to solve the equations
for the Green functions, with suitable boundary conditions.

For the wave equation:

\begin{equation}
\Box \tilde{G} (x) = \delta (x)
\end{equation}

a Fourier transformation gives:

\begin{equation}
G(p) = {\left( {\vec{p}}^{\,2} - p_0^2 \right)}^{-1} \equiv
{\left( p_{\mu} p^{\mu} \right)}^{-1} \equiv P^{-1}
\end{equation}

Of course, it is necessary to specify the nature of the singularity.
Different determinations imply different types of Green functions.
For the classical solution of (4) it is natural to use the retarded
function $ ( {\tilde{G}}_{rt} ) $. It corresponds to the propagation
towards the future of the effect produced by the sources. This
function can be obtained by means of a Fourier transform of (5)
in which the $ p_0 $ integration is taken along a path from
$ -\infty $ to $ +\infty $ , leaving the poles to the right. In
practice, we add to $ p_0 $ a small positive imaginary part:

\[G_{rt}(p) = {\left[ {\vec{p}}^{\,2} - {(p_0 + i0)}^2 \right]}^{-1} = \]
\begin{equation}
{\left( {\vec{p}}^{\,2} - p_0^2 - i 0\; sgn p_0 \right)}^{-1} =
{\left( P - i 0\; sgn p_0 \right)}^{-1}
\end{equation}

The advanced solution is the complex conjugate of (6):

\begin{equation}
G_{ad}(p) = {\left( {\vec{p}}^{\,2} - p_0^2 + i 0\; sgn p_0 \right)}^{-1} =
{\left( P + i 0\; sgn p_0 \right)}^{-1}
\end{equation}

For the Feynman propagator we have to add a small imaginary part
to $ P $ ( not just to $ p_0 $ ) :

\begin{equation}
G_{\pm}(p) = { \left( P \pm i 0 \right) }^{-1}
\end{equation}

And, in the massive case:

\begin{equation}
G_{\pm}(p) = {\left( P + m^2 \pm i 0 \right)}^{-1}
\end{equation}

The Wheeler function is half advanced and half retarded. It is easy
to see that it is also half Feynman and half its conjugate ( we will
not use any index for the Wheeler propagator):

\begin{equation}
G(p) = \frac {1} {2} G_{+}(p) + \frac {1} {2} G_{-}(p)
\end{equation}

On the real axis, the Wheeler function coincides with Cauchy's
``principal value'' Green function, which is known to be zero
on the mass-shell ( no free waves ).

We can write:

\begin{equation}
G_{\pm}(p) = G(p) \pm i \pi \delta \left( P + m^2 \right)
\end{equation}

where:

\begin{equation}
i \pi \delta \left( P + m^2 \right) = \frac {1} {2} G_{+}(p) -
\frac {1} {2} G_{-}(p)
\end{equation}

Equation (11) is a decomposition of the Feynman function into two
terms. The first one only contains virtual propagation. The second
one is a Lorentz invariant solution of the homogeneous equation
representing the free particle.

To perform convolution integrations in p-space, we will utilize
the method presented in reference \cite{tp9}. Essencially, it consists
in the use of the Bochner theorem for the reduction of the Fourier
transform to a Hankel transform. The nucleus of this transformation
is made to correspond to an arbitrary number of dimensions
$ \nu $ , taken as a free parameter. In this way, starting with a
given propagator in p-space, we get a function in x-space whose
singularity at the origin depends analytically on $ \nu $ . It exists
then a range of values ( of $ \nu $ ) such that the product of Green
functions is allowed and well determined.

In x-space we define:

\[Q = r^2 - x_0^2 = x_{\mu} x^{\mu} \]

The Fourier transform of the massless Feynman function is:

\[{\cal{F}}\{(P - i0)^{-1} \}(x) =
\frac {1} {{\left( 2 \pi \right)}^{ \frac {\nu} {2} }} \int d^{\nu}p\;
{(P -i 0)}^{-1} e^{ipx} \]

By means of a ``Wick rotation'', the $p_0$-integration can be made
to run along the imaginary axis, without crossing any pole.
Mathematically, we perform a dilatation:

\begin{equation}
p_0 = a p^{'}_0 \;\; ; \;\; x_0 = \frac {1} {a} x^{'}_0 \;\; ; \;\;
p_0 x_0 = p^{'}_0 x^{'}_0
\end{equation}

A subsequent continuation to a = i produces the transformation:

\[ P \Rightarrow {\vec{p}}^{\,2} + p^{'2}_0 = P^{'} \]
\[ Q \Rightarrow {\vec{x}}^{\,2} + x^{'2}_0 = Q^{'} \]

The new quadratic forms are Euclidean and Bochner theorem \cite{tp10}
tells that the Fourier transformation reduces to:

\begin{equation}
{\cal{F}}\left\{( P - i0 )^{-1} \right\}(x) =
\frac {i} { x^{\frac {\nu} {2}}} \int\limits_0^{\infty}dy\;
\frac { y^{ \frac {\nu} {2}}} {y^2}
{\cal{J}}_{ \frac {\nu} {2} -1} (xy)
\end{equation}

where $ {\cal{J}}_{\alpha} $ is a Bessel function of the first kind
and order $ \alpha $ .

In eq.(13) we see that a Wick rotation in p-space ($ a\rightarrow i $ )
implies an anti-Wick rotation in x-space ($ a^{-1} \rightarrow -i $ ).
We must then choose:

\[ x = {\left( Q + i0 \right)}^{\frac {1} {2}} \]

Note that the imaginary unit in eq.(14) ( r.h.s. ), is due to the
transformation $ dp_0 \rightarrow i dp^{'}_0 $

From ref.\cite{tp11} we take:

\[ \int\limits_0^{\infty} dy\; y^{\mu} {\cal{J}}_{\rho}(ay) =
2^{\mu} a^{-\mu -1}
\frac {\Gamma \left( \frac {1+\rho +\mu} {2} \right)}
{\Gamma \left( \frac {1+\rho - \mu} {2} \right) } \]

I.e.:

\begin{equation}
{\cal{F}}\left\{\left(P- i0\right)^{-1}\right\}(x) = i
2^{\frac {\nu} {2} -2} \Gamma \left( \frac {\nu} {2} -1 \right)
{\left(Q + i0\right)}^{ 1 -\frac {\nu} {2}}
\end{equation}

More general, for a function $ f(P\pm i0) $, we obtain:

\begin{equation}
{\cal{F}}\left\{f(P\pm i0)\right\}(x) = \mp
\frac {i} {x^{\frac {\nu} {2} -1}} \int\limits_0^{\infty}
dy \; y^{\frac {\nu} {2}} f\left( y^2 \right)
{\cal{J}}_{\frac {\nu} {2} -1} (xy)
\end{equation}

where $ x = (Q\mp i0)^{\frac {1} {2}} $.

The right hand side of eq.(16) is a Hankel transform of the function
$ f(y^2) $ \cite{tp12}.

\section{Massless case}

{\bf a) Fourier transforms}

With the procedures described in section 1, we can obtain the Fourier
transforms of general massless Feynman functions, defined as
$ (P\pm i0 )^{\alpha} $.

From (18) and ref.\cite{tp11}, we get ( compare with ref.\cite{tp13} ):

\begin{equation}
{\cal{F}}\left\{(P \pm i0 )^{\alpha}\right\}(x)= \mp i
2^{ 2\alpha + \frac {\nu} {2} }
\frac {\Gamma \left( \alpha + \frac {\nu} {2} \right) }
{\Gamma (-\alpha) }
\left( Q \mp i0\right)^{-\alpha -\frac {\nu} {2}}
\end{equation}

The exponent of $ Q \mp i0 $ can be deduced by dimensional
considerations. Furthermore, if we interchange the quadratic forms
$ P \leftrightarrow Q $ and write $ {\cal{F}}^{-1} $  for
$ {\cal{F}} $, then eq.(17) is still valid.

We define the massless Wheeler propagator as:

\begin{equation}
P^{\alpha} = \frac {1} {2} \left( P + i0 \right)^{\alpha} +
\frac {1} {2} \left( P - i0 \right)^{\alpha}
\end{equation}

( We will not use any index when a Wheeler function is meant )

The Fourier transform of (18) is ( cf. eq.(17) ):

\begin{equation}
{\cal{F}}\left\{P^{\alpha}\right\}(x) = i
2^{ 2\alpha + \frac {\nu} {2} }
\frac {\Gamma\left( \alpha + \frac {\nu} {2} \right) }
{\Gamma \left(-\alpha\right) }
\left[ \frac {1} {2}
\left( Q + i0 \right)^{-\alpha - \frac {\nu} {2} }
- \frac {1} {2}
\left( Q - i0\right)^{-\alpha -\frac {\nu} {2} }\right]
\end{equation}

But we also have the relation ( valid for any quadratic form
\cite{tp13})

\begin{equation}
\left( Q \pm i0 \right)^{\lambda} = Q_{+}^{\lambda} +
e^{\pm i \pi \lambda} Q_{-}^{\lambda}
\end{equation}

where

\[ Q_{+}^{\lambda} = \left\{ \begin{array}{ll}
                               Q^{\lambda} & Q>0 \\
                               0           & Q\leq 0
                             \end{array} \right. \]
\[ Q_{-}^{\lambda} = \left\{ \begin{array}{ll}
                               (-Q)^{\lambda} & Q<0 \\
                               0              & Q\geq 0
                             \end{array} \right. \]

So that we can write (19) in the form:

\begin{equation}
{\cal{F}}\left\{ P^{\alpha}\right\}(x) =
2^{2\alpha + \frac {\nu} {2}}
\frac { \Gamma \left( \alpha + \frac {\nu} {2} \right) }
{ \Gamma \left( -\alpha\right) }
sin\pi \left( \alpha + \frac {\nu} {2} \right)
Q_{-}^{-\alpha - \frac {\nu} {2} }
\end{equation}

Equation (21) shows another interesting property of Wheeler
functions. They are real and have support inside the light-cone
of the coordinates. Furthermore, for $ \alpha = -1 $, the
trigonometric function tends to zero for
$ \nu \rightarrow 4 $ , but $ Q_{-}^{1-\frac {\nu} {2}} $ has a
pole at $ \nu = 4 $ with residue $ \delta (Q) $ \cite{tp14}. Then:

\begin{equation}
\lim_{\nu \rightarrow 4} {\cal{F}}\left\{ P^{-1}\right\}(x)
= \delta (Q)
\end{equation}

In four dimensions the massless Wheeler function is concentrated
on the light cone.
From eq.(21) we obtain:

\[ {\cal{F}}^{-1}\left\{ Q_{-}^{\lambda} \right\}(p) = -
\frac { 2^{2\lambda}+\frac {\nu} {2}} {sin\pi\lambda}
\frac {\Gamma \left( \lambda + \frac {\nu} {2} \right) }
{\Gamma \left( -\lambda \right) }
P^{-\lambda - \frac {\nu} {2} } = \]
\begin{equation}
2^{2\lambda + \frac {\nu} {2} } \Gamma \left( \lambda + 1 \right)
\Gamma \left( \lambda + \frac {\nu} {2} \right)
P^{-\lambda -\frac {\nu} {2}}
\end{equation}

where the relation

\[\Gamma(z) \Gamma(1-z) = \frac {\pi} {\sin \pi z} \]

has been used.

From (17) we can also get:

\[ {\cal{F}}^{-1}\left\{Q_{+}^{\lambda}\right\}(p) = -
2^{2\lambda + \frac {\nu} {2} } \Gamma\left(\lambda + 1 \right)
\Gamma\left( \lambda + \frac {\nu} {2} \right)\;\;\times \]
\begin{equation}
\left[ cos \pi \lambda\;P_{+}^{-\lambda -\frac {\nu} {2}} +
cos\frac {\pi} {2} \nu\; P_{-}^{-\lambda -\frac {\nu} {2}} \right]
\end{equation}

From relation (12) we find:

\begin{equation}
2\pi i \delta (P) = {\left(P-i0\right)}^{-1} -
{\left(P+i0\right)}^{-1}
\end{equation}

And

\begin{equation}
{\cal{F}}\left\{ \delta (P)\right\}(x)=
\frac {2^{\frac {\nu} {2} - 2}} {2\pi}
\Gamma\left( \frac {\nu} {2} -1\right)
\left[ \left( Q+i0\right)^{1-\frac {\nu} {2}} +
\left( Q-i0\right)^{1-\frac {\nu} {2}} \right]
\end{equation}

\begin{equation}
{\cal{F}}\left\{ \delta (P)\right\}(x)=
\frac {2^{\frac {\nu} {2} - 2}} {2\pi}
\Gamma\left( \frac {\nu} {2} -1\right)
\left[  Q_{+}^{1-\frac {\nu} {2}} -
cos\frac {\pi} {2} \nu Q_{-}^{1-\frac {\nu} {2}} \right]
\end{equation}

Note that eq.(21) allows us to obtain the advanced and retarded
components of the Wheeler functions in x-space.

\[\tilde{G}_{rt}(x) = 2 \Theta (t)
{\cal{F}} \left\{ P^{\alpha} \right\}(x) \;\;\;
\tilde{G}_{ad}(x) = 2 \Theta (-t)
{\cal{F}} \left\{ P^{\alpha} \right\}(x) \]

where $ \Theta (t) $ is Heaviside's step function.

{\bf b) Convolutions}

The well-known convolution theorem asserts that the Fourier transform
of a convolution product is the product of the Fourier transform
of each factor:

\[ f(p) \ast g(p) = c {\cal{F}}^{-1}
\left\{ {\cal{F}} \left\{ f(p) \right\}(x)
{\cal{F}} \left\{ g(p) \right\}(x) \right\}(p) \]
\begin{equation}
c = {\left( 2\pi \right)}^{\frac {\nu} {2}}
\end{equation}

Formula (28) can be taken as a definition of the operation $ \ast $.
The product of distributions inside the curly brackets, can be
taken in a suitable range of $ \nu $, and analytically extended to
other values.

For Feynman propagators (28) gives:

\[\left(P-i0\right)^{\alpha} \ast \left(P-i0\right)^{\alpha} =
ic2^{-\frac {\nu} {2} }
\frac {{\Gamma}^2 \left(\alpha + \frac {\nu} {2} \right)}
{{\Gamma}^2 \left(-\alpha\right)}\;\;\times \]
\begin{equation}
\frac {\Gamma\left(-2\alpha-\frac {\nu} {2}\right)}
{\Gamma\left(2\alpha+\nu\right)}
\left(P-i0\right)^{2\alpha+\frac {\nu} {2}}
\end{equation}

An analogous equation can be obtained by changing the sign of the
imaginary unit in (29). The convolution of two Feynman functions
of the same kind (+ or -) gives another Feynman function of the same
kind.

For Wheeler propagators, eqs.(21), (23) and (28) give:

\[P^{\alpha}\ast P^{\alpha}=2^{-\frac {\nu} {2} - 1} c
\frac {{\Gamma}^2 \left(\alpha + \frac {\nu} {2} \right)}
{{\Gamma}^2 \left(-\alpha\right)} \cdot
\frac {\Gamma\left(-2\alpha-\frac {\nu} {2}\right)}
{\Gamma\left(2\alpha+\nu\right)}\;\;\times \]
\begin{equation}
tg\pi\left(\alpha + \frac {\nu} {2}\right)
P^{2\alpha + \frac {\nu} {2}}
\end{equation}

So that the convolution of two Wheeler functions gives another Wheeler
function.

For the wave equation we choose $\alpha =-1 $:

\begin{equation}
\left(P-i0\right)^{-1} \cdot \left(P-i0\right)^{-1} = 2ia(\nu)
\left(P-i0\right)^{\frac {\nu} {2} -2}
\end{equation}
\begin{equation}
\left(P+i0\right)^{-1} \cdot \left(P+i0\right)^{-1} = -2ia(\nu)
\left(P+i0\right)^{\frac {\nu} {2} -2}
\end{equation}
\begin{equation}
P^{-1}\ast P^{-1} = a(\nu)\;
tg\pi\left(\frac {\nu} {2} -1 \right) P^{\frac {\nu} {2}-2}
\end{equation}
where
\[a(\nu)=c2^{-\frac {\nu} {2} - 1}
{\Gamma}^2 \left( \frac {\nu} {2} - 1 \right)
\frac {\Gamma\left(2-\frac {\nu} {2}\right)}
{\Gamma \left( \nu - 2 \right)} \]

Eqs.(31) and (32) have a pole for $ \nu\rightarrow 4 $ (the usual
ultraviolet divergence), while (33) is well determined in that
limit.

The convolution of $\delta (P) $ function can be found with the help
of eqs.(23), (24), (27) and (28):
\begin{equation}
{\pi}^2 \delta (P) \ast \delta (P) = a(\nu)\;
tg \pi \left(\frac {\nu} {2} - 1 \right)
\left[ P_{+}^{\frac {\nu} {2} - 2} - cos \frac {\pi} {2} \nu
P_{-}^{\frac {\nu} {2} - 2 } \right]
\end{equation}

A comparison with eq.(31) shows that:

\begin{equation}
{\pi}^2 \delta(P) \ast \delta (P) = sgnP \cdot P^{-1} \ast P^{-1}
\end{equation}

This relation implies:

\[\left(P-i0\right)^{-1} \ast \left(P+i0\right)^{-1} + {\pi}^2
\delta (P) \ast \delta (P) =
2 \Theta (P)\; P^{-1} \ast P^{-1} \]

The convolution of two Feynman propagators of different kinds
has support outside the light-cone in p-space.

\section{Massive case}

A bradyon field obeys a normal Klein-Gordon equation. Its Feynman
propagator is given by eq.(9). The Wheeler function is:

\begin{equation}
\left(P+m^2\right)^{-1} =\frac {1} {2} \left(P+m^2+i0\right)^{-1} +
\frac {1} {2} \left(P+m^2-i0\right)^{-1}
\end{equation}

The on-shell $\delta$-function, solution of the homogeneous equation
is:

\begin{equation}
\delta \left(P+m^2\right) = \frac {1} {2\pi i }
\left[\left(P+m^2+i0\right)^{-1} - \left(P+m^2-i0\right)^{-1}\right]
\end{equation}

{\bf a) Fourier transforms}

To find the Fourier transform of the Feynman propagators we use
eq.(16) and ref.\cite{tp11} ( p. 687 - 6.566 - 2 ):

\[{\cal{F}}\left\{\left(P+m^2\pm i0\right)^{-1}\right\}(x) =
\mp i m^{\frac {\nu} {2} - 1}
\left(Q\mp i0\right)^{\frac {1} {2} \left( 1-\frac {\nu} {2}\right)}
{\cal{K}}_{\frac {\nu} {2}-1}
\left[m\left(Q\mp i0\right)^{\frac {1} {2}} \right] \]

where ${\cal{K}}_{\alpha}$ is a Bessel function of the third kind
and order $\alpha$. (For the definition of different Bessel functions
we follow ref.\cite{tp11}, p.951 - 8.40). Note also that with the more
general formula of ref.\cite{tp11} (p.687 - 6.565 - 4), we could
find the Fourier transform of arbitrary powers of the Feynman
propagator.

Using the relation (20), we can write:

\[{\cal{F}}\left\{\left(P+m^2\pm i0\right)^{-1}\right\}(x) =
\mp im^{\frac {\nu} {2} - 1}
\left[Q_{+}^{\frac {1} {2} \left(1-\frac {\nu} {2}\right)}
{\cal{K}}_{\frac {\nu} {2}-1}
\left(mQ_{+}^{\frac {1} {2}}\right) + \right.\]
\begin{equation}
i\frac {\pi} {2}
Q_{-}^{\frac {1} {2}\left(1-\frac {\nu} {2}\right)}
{\cal{H}}_{1-\frac {\nu} {2}}^{\beta}
\left.\left(mQ_{-}^{\frac {1} {2}}\right)  \right]
\end{equation}

where $\beta =1$ for the upper sign and $\beta =2$ for the lower
sign.

It is now easy to find the transforms of (36) and (37):

\begin{equation}
{\cal{F}}\left\{\left(P+m^2\right)^{-1}\right\}(x) = \frac {\pi} {2}
m^{\frac {\nu} {2}-1}
Q_{-}^{\frac {1} {2}\left(1-\frac {\nu} {2}\right)}
{\cal{J}}_{1-\frac {\nu} {2}}
\left(mQ_{-}^{\frac {1} {2}}\right)
\end{equation}
\[{\cal{F}}\left\{\delta \left(P+m^2\right)\right\}(x) =
-\frac {1} {\pi} m^{\frac {\nu} {2} -1}
\left[Q_{+}^{\frac {1} {2}\left(1-\frac {\nu} {2}\right)}\right.
{\cal{K}}_{\frac {\nu} {2} -1}
\left(mQ_{+}^{\frac {1} {2}}\right) - \]
\begin{equation}
\frac {\pi} {2}
Q_{-}^{\frac {1} {2}\left(1-\frac {\nu} {2}\right)}
{\cal{N}}_{1-\frac {\nu} {2} }
\left.\left(mQ_{-}^{\frac {1} {2}} \right)\right]
\end{equation}

Also for the massive case, the Wheeler function is zero outside the
light-cone.

With some sligth changes in notation, we can check that (38) and
(40) coincide with the results of ref.\cite{tp13}.

{\bf b) Convolutions}

We are going to follow the procedure already used in section 2.

For the convolution of a massive Feynman propagator with a massless
one, eqs.(17), (38) and (28) give:

\[\left(P+m^2-i0\right)^{-1} \ast \left(P-i0\right)^{-1} = -
2^{\frac {\nu} {2} -2} c
m^{\frac {\nu} {2} -1}
\Gamma\left(\frac {\nu} {2} -1\right)\;\;\times \]
\[{\cal{F}}^{-1}\left\{
\left(Q+i0\right)^{\frac {3} {2} \left(1-\frac {\nu} {2}\right) }
{\cal{K}}_{\frac {\nu} {2}-1 }
\left[m\left(Q+i0\right)^{\frac {1} {2}}\right]\right\}(p)\]

And with the help of eq.(16) and ref.\cite{tp11} (p.643 - 6.576 - 3),
we get:

\[\left(P+m^2-i0\right)^{-1} \ast \left(P-i0\right)^{-1} = i
2^{\frac {\nu} {2}} c
\frac {m^{\nu -2}} {\Gamma\left(\frac {\nu} {2}\right)}
\Gamma\left(\frac {\nu} {2}-1\right)
\Gamma\left(2-\frac {\nu} {2}\right)\;\;\times \]
\begin{equation}
F\left(1, 2-\frac {\nu} {2}, \frac {\nu} {2};
-\frac {P-i0} {m^2} \right)
\end{equation}

where $F(a, b, c; z)$ is Gauss hypergeometric function.

For two Wheeler propagators and two on-shell $\delta$-functions
we can follow the same method. It is then not difficult to prove
the relation (compare with (35)):

\[{\pi}^2 \delta\left(P+m^2\right) \ast
\delta\left(P\right) = sgnP \cdot
\left(P+m^2\right)^{-1}\ast P^{-1} \]

Or, more general:

\begin{equation}
{\pi}^2 \delta\left(P+m_1^2\right) \ast
\delta\left(P+m_2^2\right) = sgnP \cdot
\left(P+m_1^2\right)^{-1}\ast
\left(P+m_2^2\right)^{-1}
\end{equation}

\section{Tachyons}

A tachyon field obeys a Klein-Gordon equation with the wrong sign
of the ``mass'' term. The Green function is an inverse of
$P-{\mu}^2$ (we use ${\mu}^2 =-m^2$ for the mass of the tachyon).

Although it is not easy to see what a Feynman propagator should be
when the inverse of $P-{\mu}^2$=${\vec{p}}^{\,2}-p_0^2-{\mu}^2$ has
a pair of imaginary poles ($if\;\vec{p}^{\,2}<{\mu}^2$), we may nevertheless
define:

\[\left(P-{\mu}^2 \pm i0\right)^{-1} =
-\left(-P+{\mu}^2 \mp i0\right)^{-1}\]

Or, introducing the ``dual'' quadratic form:

\[{}^{+}P=-P=p_0^2-{\vec{p}}^{\,2}\]
\[\left(P-{\mu}^2\pm i0 \right)^{-1}=-
\left({}^{+}P+{\mu}^2\mp i0\right)^{-1}\]

In other words, instead of propagators with the wrong sign of the
mass, we can say that we have propagators with the wrong sign of the
metric.

To find the Fourier transform, we recall (see section 2) that we
have an ``i'' for the Wick rotation of $p_0$. For the dual metric we
have three Wick rotations. The three factors $d\vec{p}$ contribute
with $i^3=-i$. We also note that ${}^{+}Q_{+}=Q_{-}$ and
${}^{+}Q_{-}=Q_{+}$. So that (compare with (38)):

\[{\cal{F}}\left\{\left(P-{\mu}^2\pm i0\right)^{-1}\right\}(x) =
\pm i {\mu}^{\frac {\nu} {2} -1}
\left[ \frac {{\cal{K}}_{\frac {\nu} {2}-1}
\left(\mu Q_{-}^{\frac {1} {2}}\right)}
{Q_{-}^{\frac {1} {2}\left(\frac {\nu} {2}-1\right)}}
\mp \right. \]
\begin{equation}
\left. i\frac {\pi} {2} \frac {{\cal{H}}^{\beta}_{1-\frac {\nu} {2}}
\left(\mu Q_{+}^{\frac {1} {2}}\right)}
{Q_{+}^{\frac {1} {2}\left(\frac {\nu} {2}-1\right)}}\right]
\end{equation}

where $\beta =2$ for the upper sign and $\beta=1$ for the lower sign.

The real part of (43) is:

\[\frac {1} {2} {\cal{F}}
\left\{\left(P-{\mu}^2+i0\right)^{-1} \right\}(x) +
\frac {1} {2} {\cal{F}}
\left\{\left(P-{\mu}^2-i0\right)^{-1} \right\}(x) =\]
\begin{equation}
Re{\cal{F}}\left\{\left(P-{\mu}^2\pm i0\right)^{-1} \right\}(x)= 
\frac {\pi} {2} {\mu}^{\frac {\nu} {2}-1}
Q_{+}^{\frac {1} {2} \left(1-\frac {\nu} {2}\right)}
{\cal{J}}_{1-\frac {\nu} {2}}
\left(\mu Q_{+}\right)
\end{equation}

This real part has support outside the light-cone, while for bradyons,
the real part of the Feynman propagator is zero for $x^{\mu}$
space-like (cf. eq.(39)).

We will now show that (44) is not the Wheeler propagator for the
tachyon. To this aim, we go back to the original definition.
Namely, a half retarded and half advanced propagator.

\begin{equation}
\left(P-{\mu}^2\right)^{-1}=\frac {1} {2}
\left(P-{\mu}^2\right)_{Ad}^{-1} + \frac {1} {2}
\left(P-{\mu}^2\right)_{Rt}^{-1}
\end{equation}

The Fourier transform is:

\[{\cal{F}}\left\{\left(P-{\mu}^2\right)^{-1}\right\}(x) =
\frac {1} {\left(2\pi\right)^{\nu /2}}
\frac {1} {2}
\int d^{\nu}p \frac {e^{ipx}} {\left(P-{\mu}^2\right)_{Ad}} + \]
\begin{equation}
\frac {1} {\left(2\pi\right)^{\nu /2}}
\frac {1} {2}
\int d^{\nu}p \frac {e^{ipx}} {\left(P-{\mu}^2\right)_{Rt}}
\end{equation}

We will first evaluate the advanced part of (46):

\begin{equation}
{\cal{F}}\left\{\left(P-{\mu}^2\right)_{Ad}^{-1}\right\}(x) =
\frac {1} {\left(2\pi\right)^{\nu /2}} \int d^{\nu-1}p\;
e^{i\vec{p}\cdot\vec{r}}
\int\limits_{Ad} dp_0
\frac {e^{-ip_0x_0}} {{\vec{p}}^{\,2}-p_0^2-{\mu}^2}
\end{equation}

Where the path of integration runs parallel to the real axis and
below both poles of the integrand. For $x_0<0$ the path can be closed
on the upper half plane of $p_0$. The contribution will be the
residues at the poles:

\[p_0=\pm\omega=\pm\sqrt{{\vec{p}}^{\,2}-{\mu}^2}\;\;\;,if\;\;\;
{\vec{p}}^{\,2}\geq {\mu}^2 \]
\[p_0=\pm i{\omega}^{'}=\pm i\sqrt{{\mu}^2-{\vec{p}}^{\,2}}\;\;\;,if\;\;\;
{\vec{p}}^{\,2} \leq {\mu}^2 \]
\[{\cal{F}}\left\{\left(P-{\mu}^2\right)_{Ad}^{-1}\right\}(x)=
\frac {1} {\left(2\pi\right)^{\nu /2}} \int d^{\nu-1}p
e^{i\vec{p}\cdot\vec{r}} 2\pi i
\left[\left(\frac {e^{i\omega x_0}} {2\omega} -
\frac {e^{-i\omega x_0}} {2\omega}\right)
\Theta\left({\vec{p}}^{\,2}-{\mu}^2\right) \right.\]
\[+ \left.\left(\frac {-e^{\omega^{'} x_0}} {2i\omega^{'}} +
\frac {e^{-\omega^{'} x_0}} {2i\omega^{'}}\right)
\Theta\left({\mu}^2-{\vec{p}}^{\,2}\right) \right] = \]
\begin{equation}
\frac {-2\pi} {\left(2\pi\right)^{\nu /2}}
\int d^{\nu-1}p e^{i\vec{p}\cdot\vec{r}}
\left[\frac {sin\omega x_0} {\omega}
\Theta\left(\vec{p}^{\,2}-{\mu}^2\right) +
\frac {sh{\omega}^{'} x_0} {{\omega}^{'}}
\Theta\left({\mu}^2-{\vec{p}}^{\,2}\right) \right]
\end{equation}

Note that we can write the brackets as (cf. eq. (20))

\[\left[\frac {sin\omega x_0} {\omega}
\Theta\left(\vec{p}^{\,2}-{\mu}^2\right) +
\frac {sh{\omega}^{'} x_0} {{\omega}^{'}}
\Theta\left({\mu}^2-{\vec{p}}^{\,2}\right) \right] = \]
\begin{equation}
\frac {sin\left[x_0 \left({\vec{p}}^{\,2}-{\mu}^2+i0\right)^{\frac {1} {2}}    
\right]}
{\left({\vec{p}}^{\,2}-{\mu}^2+i0\right)^{\frac {1} {2}}}=
\frac {sinx_0\Omega} {\Omega}
\end{equation}

For the spatial Fourier transform (48), we use Bochner theorem
(cf. eq.14))

\[{\cal{F}}\left\{\left(P-{\mu}^2\right)_{Ad}^{-1}\right\}(x) =
\left(2\pi\right)^{\frac {1} {2}}
\frac {\Theta(-t)} {r^{\frac {\nu -3} {2}}}
\int\limits_0^{\infty} dk\;k^{\frac {\nu-1} {2}}
\frac {sin\mid t\mid \Omega} {\Omega}
{\cal{J}}_{\frac {\nu -3} {2}} (rk) \]

Formula 6.737-5, p.761 of the table of ref.\cite{tp11} gives
($b=i\mu +0$)

\begin{equation}
{\cal{F}}\left\{\left(P-{\mu}^2\right)_{Ad}^{-1}\right\}(x) =
\pi {\mu}^{\frac {\nu} {2} -1}
Q_{-}^{\frac {1} {2} \left(1-\frac {\nu} {2}\right)}
{\cal{I}}_{1-\frac {\nu} {2}}
\left(\mu Q_{-}^{\frac {1} {2}}\right)\;\;\;(x_0<0)
\end{equation}

And of course:

\[{\cal{F}}\left\{\left(P-{\mu}^2\right)_{Ad}^{-1}\right\}(x) = 0\;\;\;
for\;x_0 \geq 0 \]

The retarded Fourier transform reproduces (48), with a change of
sign, for $x_0>0$ and is zero for $x_0\leq 0 $. We then get, for
the Wheeler Green function (46):

\begin{equation}
{\cal{F}}\left\{\left(P-{\mu}^2\right)^{-1}\right\}(x) =
\frac {\pi} {2} {\mu}^{\frac {\nu} {2} -1}
Q_{-}^{\frac {1} {2} \left(1-\frac {\nu} {2}\right)}
{\cal{I}}_{1-\frac {\nu} {2}}
\left(\mu Q_{-}^{\frac {1} {2}}\right)
\end{equation}

Again, the Wheeler propagator has support inside the light-cone.
But, instead of a Bessel function of the first kind, we have now
a Bessel function of the second kind.

It is clear that by taking into account eq.(28), we can evaluate
convolutions of different Green functions. As it was done in
section 3 for bradyons fields.

\section{Fields with complex mass parameters}

The descomposition in Klein-Gordon factors of a higher order
equation, often leads to complex mass parameters. Equation (1)
is an example. The constituent fields obey eq.(2). A simple higher
order equation such as (3) presents the same behaviour. Of course
for a real equation the masses come in complex conjugate pairs.
We consider:

\begin{equation}
\left(\Box-M^2\right)\phi = 0\;\;,\;\;M=m+i\mu\;\;\;(m>0)
\end{equation}

This type of equation has been analyzed elsewhere (ref.\cite{tp5}).
The Green functions are inverses of $P+M^2$ =${\Omega}^2-p_0^2 $,
where $\Omega$=$({\vec{p}}^{\,2}+M^2)^{1/2}$. The two poles at
$p_0\pm\Omega$, move when $\vec{p}^{\,2}$ varies from 0 to $\infty$, on a
line contained in a band of width $\pm i\mu$, centered at the real
axis.

The retarded Green function is obtained with a $p_0$-integration that
runs parallel to the real axis, with $Imp_0>|\mu|$. For the advanced
solution, the integration runs below both poles ($Imp_0<-|\mu|$):

\[{\cal{F}}\left\{\left(P+M^2\right)_{Ad}^{-1}\right\}(x)=
\frac {1} {\left(2\pi\right)^{\nu/2}} \int d^{\nu-1}p\;
e^{i\vec{p}\cdot\vec{r}} \int\limits_{Ad} dp_0 \;
\frac {e^{ip_0t}} {{\Omega}^2-p_0^2} =\]
\[\frac {1} {\left(2\pi\right)^{\nu/2}} \int d^{\nu-1}p\;
e^{i\vec{p}\cdot\vec{r}}
\frac {2\pi i \Theta(-t)} {2\Omega}
\left(e^{i\Omega t} - e^{-i\Omega t}\right)= \]
\[-2\pi \frac {\Theta(-t)} {\left(2\pi\right)^{\nu/2}}
\int d^{\nu-1}p\; e^{i\vec{p}\cdot\vec{r}}
\frac {sin\Omega t} {\Omega}\]
\[ {\cal{F}}\left\{\left(P+M^2\right)_{Ad}^{-1}\right\}(x)=
\left(2\pi\right)^{\frac {1} {2}}
\frac {\Theta(-t)} {r^{\frac {\nu-3} {2}}}
\int\limits_0^{\infty} dk\; K^{\frac {\nu-1} {2}}
\frac {sin\Omega\mid t\mid} {\Omega}
{\cal{J}}_{\frac {\nu-3} {2}} (rk) \]

And according to ref.\cite{tp11} (6.735 - 5):

\[ {\cal{F}}\left\{\left(P+M^2\right)_{Ad}^{-1}\right\}(x)=
\pi\Theta(-t) M^{\frac {\nu} {2} -1}
Q_{-}^{\frac {1} {2}\left(1-\frac {\nu} {2}\right)}
{\cal{J}}_{\frac {\nu-3} {2}}
\left(MQ_{-}^{\frac {1} {2}}\right) \]

For the retarded Green function we get the same answer, with
$\Theta(+t)$ replacing $\Theta(-t)$.

The Wheeler propagator is then:

\begin{equation}
{\cal{F}}\left\{\left(P+M^2\right)^{-1}\right\}(x)=
\frac {\pi} {2} M^{\frac {\nu} {2} -1}
Q_{-}^{\frac {1} {2}\left(1-\frac {\nu} {2}\right)}
{\cal{J}}_{\frac {\nu-3} {2}}
\left(MQ_{-}^{\frac {1} {2}}\right)
\end{equation}

Now we have the general result: The Wheeler function propagates
inside the light-cone for any value of the mass, real (bradyons),
imaginary (tachyons) or complex ($M=m+i\mu$).

In the case of complex masses, a natural definition for the Feynman
propagator is obtained by a $p_0$-integration along the real axis. It
is not difficult to see that

\[{\cal{F}}\left\{\left( P+M^2\right)_F^{-1}\right\}(x) =
\sqrt{\frac {\pi} {2}}  r^{\frac {3-\nu} {2}}
\int\limits_0^{\infty} dk\; k^{\frac {\nu -1} {2}}
\left(\frac {sin\Omega\mid t\mid} {\Omega}\right. - \]
\begin{equation}
\left. i\;sgn\mu\;
\frac {cos\Omega\mid t\mid} {\Omega}\right)
{\cal{J}}_{\frac {\nu -3} {2}}(rk)
\end{equation}

The first term in the right hand side is the Wheeler function.
The second term corresponds to a positive loop around the pole
in the upper half-plane, and a negative loop around the pole
in the lower half-plane.

The conjugate Feynman function ({\bf not} the complex conjugate
one ), is obtained by changing the sign of the second term. I.e.:
by changing the sign of both loops.

The term in $cos\Omega |t| $ can be read in ref.\cite{tp11}
(6.735 - 6). With these definitions the Wheeler function is
half Feynman and half its conjugate.

Note that for a real mass, $\mu$ corresponds to the ``small''
negative imaginary part added to the mass in Feynman's definition.
Accordingly, for real mass we take $sgn\mu =-1$.

Equation (54) can be utilized to define:

\begin{equation}
{\cal{F}}\left\{\delta\left(P+M^2\right)\right\}(x) = -
\frac {sgn\mu} {\pi} \sqrt{\frac {\pi} {2}}
r^{\frac {3-\nu} {2}}
\int\limits_0^{\infty} dk\; k^{\frac {\nu -1} {2}}
\frac {cos\Omega\mid t\mid} {\Omega}
{\cal{J}}_{\frac {\nu -3} {2}}(rk)
\end{equation}
\begin{equation}
\left(P+M^2\right)^{-1}_F = \left(P+M^2\right)^{-1} + i \pi
\delta\left(P+M^2\right)
\end{equation}

The last formula is valid for any mass, real or complex.

\section{Associated vacuum}

It is well known that the perturbative solution to the quantum
equation of motion leads to a Green function which is the vacuum
expectation value of the chronological product of field operators
(VEV). Furthermore, when the quanta are not allowed to have negative
energies, the VEV turns out to be Feynman's propagator.

However, when the energy-momentum vector is space-like the sign of
its energy component is not Lorentz invariant. It is then natural
to have symmetry between positive and negative energies. It has
been shown in references \cite{tp7} and \cite{tp8} that under this
premise, the VEV is a Wheeler propagator.

To see clearly the origin of the difference between both types of
propagators, we are going to compare the usual procedure with the
symmetric one.

A quantized Klein-Gordon field can be written as:

\begin{equation}
\varphi (x)= \frac {1} {\left(2\pi\right)^{3/2}}
\int\frac {d^3k} {\sqrt{2\omega}} \left[ a(\vec{k})\; e^{ik\cdot x} +
a^{+}(\vec{k})\; e^{-ik\cdot x} \right]
\end{equation}

where

\[\left[a(\vec{k}), a^{+}({\vec{k}}^{'})\right] =
\delta (\vec{k}-{\vec{k}}^{'})\;\;\;;\;\;\;
\omega=\sqrt{{\vec{k}}^{\,2} + m^2} \]

For simplicity, we are going to consider a single (discretized)
degree of freedom.

The raising and lowering operators obey:

\begin{equation}
\left[a, a^{+}\right]=1
\end{equation}

The energy operator is:

\[h=\frac {\omega} {2} \left(aa^{+}+a^{+}a\right)=
\omega a^{+}a+\frac {\omega} {2} =
h_0+\frac {\omega} {2} \]

Usually, the energy is redefined to be $h_0$. The vacuum then obeys:

\begin{equation}
h_0\mid 0>=0
\end{equation}

It is a consequence of (58) and (59) that:

\begin{equation}
<0|aa^{+}|0>=1\;\;\;,\;\;\;<0|a^{+}a|0>=0
\end{equation}

On the other hand, the symmetric vacuum is defined to be the state
that has zero ``true energy'':

\begin{equation}
h|0>=\frac {\omega} {2} \left(aa^{+}+a^{+}a\right) |0>=0
\end{equation}

Equations (58) and (61) imply:

\begin{equation}
<0|aa^{+}|0>=\frac {1} {2}\;\;,\;\;<0|a^{+}a|0>=
-\frac {1} {2}
\end{equation}

Let as assume, for the sake of the argument, that we define a
``ceiling'' state (as opossed to a ground state):

\begin{equation}
a^{+}|0>=0
\end{equation}

Equations (58) and (63) give:

\begin{equation}
<0|aa^{+}|0>=0\;\;,\;\;<0|a^{+}a|0>=-1
\end{equation}

The usual normal case, eq.(60) leads to the Feynman propagator. The
``inverted'' case, eq.(64), leads to its complex conjugate. Then
eq.(62), which is one half of (60) and one half of (64), leads to
one half of the Feynman function and one half of its conjugate.
This is the Wheeler propagator defined in section 1.

The space of states generated by succesive aplications of $a$ and
$a^{+}$ on the symmetric vacuum has an indefinite metric.

The scalar product can be defined by means of the holomorphic
representation \cite{tp14}. The functional space is formed by
analytic functions $f(z)$, with the scalar product:

\begin{equation}
\langle f, g\rangle= \int dz\;d\overline{z}
e^{-z\overline{z}} f(z) \;\overline{g(z)}
\end{equation}

Or, in polar coordinates:

\begin{equation}
\langle f, g\rangle=\int\limits_0^{\infty} d\rho\;\rho
e^{-{\rho}^2} \int\limits_0^{2\pi} d\phi f(z)
\overline{g(z)}
\end{equation}

The raising and lowering operators are represented by:

\begin{equation}
a^{+} =z\;\;\;,\;\;\;a=\frac {d} {dz}
\end{equation}

The symmetric vacuum obeys:

\[\left( \frac {d} {dz}\;z+z\;\frac {d} {dz}\right)f_0=
\left( 1+2z\;\frac {d} {dz} \right)f_0=0 \]

whose normalized solution is:

\[f_0=\left(2{\pi}^{3/2}\right)^{-1/2} z^{-1/2} \]

The energy eigenfunctions are:

\begin{equation}
f_n =\left[2\pi\Gamma\left(n+\frac {1} {2}\right)\right]^{-1/2}
z^{-1/2}z^n
\end{equation}

\section{Unitarity}

In QFT, the equations of motions for the states of a system of
interacting fields are formally solved by means of the evolution
operator.

\[U\left(t, t_0\right)|t_0>=|t> \]

The interactions between the quanta of the fields is supossed to take
place in a limited region of space-time. The initial and final
times can be taken to be $t_0\rightarrow -\infty$ and
$t\rightarrow +\infty$. Thus defining the S-operator:

\[S=U\left(+\infty , -\infty\right)\]

We do not intend to discuss the possible problems of such a definition.  
Here we are only interested in its relation to Wheeler propagators.

Usually, the initial and final states are represented by free particles.
However, when Wheeler fields are present, their quanta either mediate 
interactions between other particles, or they end up at an absorber.
This circunstance had been pointed out by J.A.Wheeler and R.P.Feynman
in references \cite{tp1} and \cite{tp3}. In consequence, the
S-matrix not only contains the initial and final free particles, but
it also contains the states of the absorbers. Through the latter we
can determine the physical quantum numbers of the Wheeler virtual
``asymptotic particles''. For these reasons, even if the initial and
final states do not contain any Wheeler free particle, for the 
verification of perturbative unitarity it is necessary to take them
into account.

We shall ilustrate this point with some examples. Let us consider
four scalar fields ${\phi}_s$ (s=1,...,4) obeying Klein-Gordon equations
with mass parameters $m_s^2$ and the interaction $\Lambda = \lambda
{\phi}_1{\phi}_2{\phi}_3{\phi}_4$. They can be written as in eq.(57).

Unitarity implies:

\[SS^{+}=1\]

or, with $S=1-T$:

\[T+T^{+}=TT^{+}\]

We introduce the initial and final states and also a complete 
descomposition of the unit operator:

\[<\alpha |T+T^{+}|\beta>= \int d{\sigma}_{\gamma}
<\alpha |T|\gamma><\gamma |T^+|\beta >\]

For the perturbative development:

\[T= \sum\limits_n {\lambda}^n T_n \]
\begin{equation}
<\alpha |T_n+T_n^+|\beta >=\sum\limits_{s=1}^{n-1} \int
d{\sigma}_{\gamma} <\alpha |T_{n-s}|\gamma > <\gamma |T_s^+|\beta>
\end{equation}

In particular, $T_0=0$ and $T_1$=pure imaginary.

For n=2

\begin{equation}
<\alpha |T_2+T_2^{+} |\beta >=\int d{\sigma}_{\gamma} 
<\alpha |T_1|\gamma > <\gamma|T_1^{+}|\beta >
\end{equation}

where we will take $T_1=i{\phi}_1{\phi}_2{\phi}_3{\phi}_4$.

${\phi}_1$ and ${\phi}_2$ are supposed to be normal fields whose
states can be obtained from the usual vacuum.

\[|\alpha>=a_2^{+}a_1^{+}|0>\;\;,\;\;
|\beta>= a_{2^{'}}^{+}a_{1^{'}}^{+}|0>\]

On the other hand, for ${\phi}_3$ and ${\phi}_4$ we leave open the 
posibility of a choice between the usual vacuum or the symmetric one.

The left hand side of (70) comes from the second order loop formed with 
the convolution of a propagator for ${\phi}_3$ and another for
${\phi}_4$. When both fields are normal, we have the convolution 
of two Feynman propagators, where the real part is:

\[Re\left[\left(P+m_3^2-i0\right)^{-1} \ast
\left(P+m_4^2-i0\right)^{-1}\right]=\left(P+m_3^2\right)^{-1}
\ast \left(P+m_4^2\right) - \]
\[{\pi}^2 \delta\left(P+m_3^2\right) \ast 
\delta \left(P+m_4^2\right) \]

And according to (42), we have in the physical region (P negative):

\[Re\left[\left(P+m_3^2-i0\right)^{-1} \ast
\left(P+m_4^2-io\right)^{-1}\right]=
2\left(P+m_3^2\right)^{-1} \ast \left(P+m_4^2\right)^{-1}\]
\begin{equation}
\left(P<0\right)
\end{equation}

Equation (71) implies that the left hand side of (70) for Feynman
particles is twice the value corresponding to Wheeler particles.

The relation (70) is known to be valid for normal fields. So, there
is no point in proving each. We are going to show where the relative 
factor 2 comes from.

The decomposition of unity for normal fields is:

\[{\bf I}=\int d{\sigma}_{\gamma}\; |\gamma><\gamma|= 
|0><0| + \int d^{\nu -1}q;
a^{+}\left(\vec{q}\right)|0><0|a\left(\vec{q}\right) +\]
\begin{equation}
\int d^{\nu -1}q_1\;d^{\nu-1}q_2 \frac {1} {\sqrt{2}} 
a^{+}\left({\vec{q}}_1\right)a{+}\left({\vec{q}}_2\right)|0>
<0|a\left({\vec{q}}_1\right) a\left({\vec{q}}_2\right)  
\frac {1} {\sqrt{2}}+....
\end{equation}

Then, for the $T_1$ matrix we have:

\[<\alpha|T_1|\gamma>= <0|a_1\left({\vec{p}}\right){\phi}_1(x)|0>
<0|a_2\left({\vec{p}^{'}}\right){\phi}_2(x)|0>\;\;\times\]
\begin{equation}
<0|{\phi}_3(x) a_3^{+}\left({\vec{q}}_3\right)|0>
<0|{\phi}_4(x) a_4^{+}\left({\vec{q}}_4\right)|0>
\end{equation}

where an integration over x-space is understood.

When the fields are expressed in terms of the operators $a(q)$ and
$a^{+}(q)$, as in equation (59) we obtain:

\begin{equation}
<\alpha|T_1|\gamma>= \frac {\left(2\pi\right)^{\nu}}
{\left(2\pi\right)^{2\left(\nu-1\right)}}
\frac {\delta\left(p-q_3-q_4\right)}{4\sqrt{{\omega}_1{\omega}_2
{\omega}_3{\omega}_4}}\;\;\;(p=p_1+p_2)
\end{equation}

And 

\begin{equation}
<\gamma|T_1|\beta>= \frac {\left(2\pi\right)^{\nu}}
{\left(2\pi\right)^{2\left(\nu-1\right)}}
\frac {\delta\left(q_3+q_4-p^{'}\right)}{4\sqrt{{\omega}_1^{'}
{\omega}_2^{'}{\omega}_3{\omega}_4}}\;\;\;
(p^{'}=p_1^{'}+p_2^{'})
\end{equation}

Multiplying together (74) and (75) and adding all possible
$|\gamma><\gamma|$ (all ${\vec{q}}_3$ and ${\vec{q}}_4$), we get:

\[\int d{\sigma}_{\gamma}<\alpha|T_1|\gamma><\gamma|T_1^{+}|\beta>=
\frac {\delta\left(p-p^{'}\right)} {16 \left(2\pi\right)^{2\nu-4}
\sqrt{{\omega}_1{\omega}_2{\omega}_1^{'}{\omega}_2^{'}}}\]
\begin{equation}
\int d\vec{q}\;\frac {\delta\left(p^0-{\omega}_3\left(\vec{q}\right)-
{\omega}_4\left(\vec{p}-\vec{q}\right)\right)}
{{\omega}_3\left(\vec{q}\right)
{\omega}_4\left(\vec{p}-\vec{q}\right)}
\end{equation}

This result coincides with (70) (l.h.s.) when the $p^0$-convolution is
carried out.

Suposse now that one of the fields, says ${\phi}_4$, has the Wheeler 
function as propagator. Instead of eq.(71) we have:

\[ Re\left[\left(P+m_3^2-i0\right)^{-1} \ast
\left(P+m_4^2-io\right)^{-1}\right]=\]
\begin{equation}
\left(P+m_3^2\right)^{-1} \ast \left(P+m_4^2\right)^{-1}
\end{equation}

Half the value of (71).

To evaluate the matrix $<T_1>$ for this case we note that the 
descomposition of unity for the states of ${\phi}_4$ (with an indefinite
metric) is now:

\[{\bf I}=\int d{\sigma}_{\gamma}\; |\gamma><\gamma|= 
|0><0| + \int d^{\nu-1}q\;
\sqrt{2} a^{+}\left(\vec{q}\right)|0><0|a\left(\vec{q}\right)\sqrt{2}-\]
\begin{equation}
\int d^{\nu-1}q\;\sqrt{2} a\left(\vec{q}\right)|0>
<0|a^{+}\left(\vec{q}\right)\sqrt{2} +....
\end{equation}

The normalization factors come from the VEV quoted in section 6, eq.(62).
It is not necessary to evaluate again the matrix element (73). Its last
vacuum expectation value has now a factor 1/2 from eq.(62), and a 
factor $\sqrt{2}$ form the normalization in (78). When the matrix for 
$T_1$ and $T_1^{+}$ are multiplied together, we get an extra factor
$(\sqrt{2}/2)^2$= $1/2$ As it should be for unitarity to hold.

When both fields ${\phi}_3$ and ${\phi}_4$, are of the Wheeler type,
we get for the convolution of the respective Wheeler propagators the
same result (77).

The matrix element of $T_1$ gains now two factors $\sqrt{2}/2$, i.e.
a factor 1/2. When we multiply $<T_1><T_1^{+}>$ we get a factor
$1/2 \cdot 1/2$= $1/4$. And we seem to be in trouble with unitarity.
However, in this case a new matrix contributes to $<T_1>$. It is:

\[<0|a_1\left(\vec{p}_1\right){\phi}_1(x)a_1^{+}\left(\vec{q}_1\right) 
a_1^{+}\left(\vec{q}_1^{\,'}\right)|0>
<0|a_2\left(\vec{p}_2\right){\phi}_2(x)a_2^{+}\left(\vec{q}_2\right) 
a_2^{+}\left(\vec{q}_2^{\,'}\right)|0>\]
\begin{equation}
<0|{\phi}_3(x)a_3^{+}\left(\vec{q}_3\right)|0>
<0|{\phi}_4(x)a_4^{+}\left(\vec{q}_4\right)|0>
\end{equation}

(79) is only possible when both, ${\phi}_3$ and ${\phi}_4$ are associated
with symmetric vacua.

For the first matrix factor we have:

\[<0|a_1\left(\vec{p}_1\right){\phi}_1(x)a_1^{+}\left(\vec{q}_1\right) 
a_1^{+}\left(\vec{q}_1^{\,'}\right)|0>=\]
\begin{equation}
\delta\left(p_1-q_1\right)
\frac {e^{-iq_1^{'}x}} {\sqrt{2{\omega}_1\left(q_1^{'}\right)}}+
\delta\left(p_1-q_1^{'}\right)
\frac {e^{-iq_1x}} {\sqrt{2{\omega}_1\left(q_1\right)}}
\end{equation}

A similar matrix factor from $<T_1^{+}>$ gives:

\[<0|a_1\left(\vec{q}_1\right)a_1\left(\vec{q}_1^{\,'}\right){\phi}_1(y) 
a_1\left(\vec{p}_1^{\,'}\right)|0>=\]
\begin{equation}
\delta\left(p_1^{'}-q_1^{'}\right)
\frac {e^{iq_1y}} {\sqrt{2{\omega}_1\left(q_1\right)}}+
\delta\left(p_1^{'}-q_1\right)
\frac {e^{iq_1^{'}y}} {\sqrt{2{\omega}_1\left(q_1^{'}\right)}}
\end{equation}

When we multiply together (80) and (81), the crossed terms do not 
contribute ($\delta(p_1-p_1^{'})=0$). The other two terms give equal
contributions. A similar evaluation can be done for the second factor
of (79) and the corresponding factor of $<T^{+}>$. For this reason
we are going to keep only the first terms from (80) and (81) 
(multiplied with the appropriate constants):

\[<\alpha|T_1|\gamma>=\frac {2}
{\left(2\pi\right)^{2\left(\nu-1\right)}}
\delta\left(p_1-q_1\right)
\frac {e^{-iq_1^{�}x}} {\sqrt{2{\omega}_1\left(q_1^{�}\right)}}
\delta\left(p_2-q_2\right)\;\;\times \]
\[\frac {e^{-iq_2^{�}x}} {\sqrt{2{\omega}_2\left(q_2^{�}\right)}}
\frac {e^{iq_3x}} {2\sqrt{2{\omega}_3\left(q_3\right)}}
\frac {e^{iq_4x}} {2\sqrt{2{\omega}_4\left(q_4\right)}}\]

And after performing the x-integration,

\[<\alpha|T_1|\gamma>=\frac {\left(2\pi\right)^{\nu}}
{2\left(2\pi\right)^{2\left(\nu-1\right)}}
\frac {\delta\left(-q_1^{'}-q_2^{'}+q_3+q_4\right)                                                                                                           
\delta\left(p_1-q_1\right) \delta\left(p_2-q_2\right)} 
{4\sqrt{{\omega}_1^{'}{\omega}_2^{'}{\omega}_3 {\omega}_4}} \]

Analogously:

\[<\gamma|T_1^{+}|\beta>=\frac {\left(2\pi\right)^{\nu}}
{2\left(2\pi\right)^{2\left(\nu-1\right)}}
\frac {\delta\left(q_1+q_2-q_3-q_4\right)                                                                                                           
\delta\left(p_1^{'}-q_1^{'}\right) \delta\left(p_2^{'}-q_2^{'}\right)} 
{4\sqrt{{\omega}_1^{'}{\omega}_2^{'}{\omega}_3 {\omega}_4}} \]

The sum $\int d\sigma_{\gamma}\;<\alpha|T_1|\gamma>
<\gamma|T_1^{+}|\beta>$ corresponds to an integration on
$\vec{q}_1$,$\vec{q}_1^{\,'}$,$\vec{q}_2$,$\vec{q}_2^{\,'}$. It is easy to 
see that after this operations we get one fourth of (76). Thus completing
the proof of unitarity, for the proposed example.

Let us now consider the case in which ${\phi}_3$ and ${\phi}_4$ are
fields obeying Klein-Gordon equations with complex mass parameters
(See section 5).

The solution of eq.(52):

\[\left(\Box -M^2\right)\phi =0 \;\;,\;\;M=m+i\mu \;\;\;(m>0) \]

can be written as:

\begin{equation}
\phi(x)=\frac {1} {\left(2\pi\right)^{\frac {\nu-1} {2}}}
\int d^{\nu-1}p\;\frac {e^{i\vec{p}\cdot\vec{x}}}
{\sqrt{2\Omega}} \left[a\left(\vec{p}\right)e^{-i\Omega t}+
b\left(\vec{p}\right)e^{i\Omega t} \right]
\end{equation}

where 

\begin{equation}
\Omega =\left({\vec{p}}^{\,2}+M^2\right)^{\frac {1} {2}}
\end{equation}

The quantization of $\phi$ leads to the rules (ref.\cite{tp5}):

\[\left[a\left(\vec{p}\right), b\left(\vec{p}^{\,'}\right)\right]=
\delta\left(\vec{p}-\vec{p}^{\,'}\right)\]

and to the adoption of the symmetric vacuum. From which we get:

\begin{equation}
<0|a\left(\vec{p}\right)b\left(\vec{p}^{\,'}\right)|0>=-
<0|b\left(\vec{p}\right)a\left(\vec{p}^{\,'}\right)|0>=
\frac {1} {2} \delta\left(\vec{p}-\vec{p}^{\,'}\right)
\end{equation}

Accordingly, the descomposition of unity for complex mass fields is
(compare with (78)):

\[{\bf I}=|0><0| + \int d^{\nu-1}q\;\sqrt{2} b\left(\vec{q}\right)|0>
<0|a\left(\vec{q}\right)|0>\sqrt{2} - \]
\[\int d^{\nu-1}q\;\sqrt{2} a\left(\vec{q}\right)|0>
<0|b\left(\vec{q}\right)|0>\sqrt{2} +.... \]

The matrix $<\alpha|T_1|\gamma>$ have the form given in eq.(73),
but now, instead of $a_3^{+}$ and $a_4^{+}$, we have to write the four
possible operators:$a_3, a_4$; $a_3, b_4$; $b_3, a_4$ and $b_3, b_4$. 

When multiplied with $<\gamma|T_1^{+}|\beta>$ as in (76) they give
similar contributions except for the signs of $\omega_3$ and
$\omega_4$ in the arguments of the $\delta$-functions. Then, the
$\delta$-function for each of the terms in the integral of (76) (r.h.s),
should be:$\delta(p_0-{\Omega}_3-{\Omega}_4)$, 
$\delta(p_0-{\Omega}_3+{\Omega}_4)$,
$\delta(p_0+{\Omega}_3-{\Omega}_4)$ and
$\delta(p_0+{\Omega}_3+{\Omega}_4)$, respectively.

This is exactly one half of the convolution product of the Wheeler
propagators for ${\phi}_3$ and ${\phi}_4$. The other half comes from
the contribution of the matrix elements of the form (79).

Of course, for every field corresponding to a complex mass M, there is
another field corresponding to the complex conjugate mass $M^{\ast}$.

We have:

\[{\phi}^{+}(x)=\frac {1} {\left(2\pi\right)^{\frac {\nu-1} {2}}}
\int d^{\nu-1}p\;\frac {e^{-i\vec{p}\cdot\vec{x}}}
{\sqrt{2{\Omega}^{\ast}}} \left[b^{+}\left(\vec{p}\right)
e^{-i{\Omega}^{\ast} t} +
a^{+}\left(\vec{p}\right)e^{i{\Omega}^{\ast} t} \right]\]
\[\left[b^{+}\left(\vec{p}\right), a^{+}\left(\vec{p}^{\,'}\right)\right]=
\delta\left(\vec{p}-\vec{p}^{\,'}\right)\]
\[<0|b^{+}\left(\vec{p}\right)a^{+}\left(\vec{p}^{\,'}\right)|0>=-
<0|a^{+}\left(\vec{p}\right)b^{+}\left(\vec{p}^{\,'}\right)|0>=
\frac {1} {2} \delta\left(\vec{p}-\vec{p}^{\,'}\right)\]
\[{\bf I}=|0><0| + \int d^{\nu-1}q\;\sqrt{2} a^{+}\left(\vec{q}\right)|0>
<0|b^{+}\left(\vec{q}\right)|0>\sqrt{2} - \]
\[\int d^{\nu-1}q\;\sqrt{2} b^{+}\left(\vec{q}\right)|0>
<0|a^{+}\left(\vec{q}\right)|0>\sqrt{2} +.... \]

The T-matrix can be constructed with the four possible contributions:
${\phi}_3$,${\phi}_4$; ${\phi}_3$, ${\phi}_4^{+}$;
${\phi}_3^{+}$, ${\phi}_4$ and ${\phi}_3^{+}$, ${\phi}_4^{+}$. 

Correspondingly, for each Wheeler propagator with mass M, there is
another one with mass $M^{\ast}$. And of course, four possible
convolution products. It is easy to see that the total convolution
is real, as well as the total T-matrix.

The proof of unitarity for the chosen example, where the masses are
complex numbers, contains as particular cases all fields with 
symmetric vacua. For bradyons M=m. For tachyons, the limit
$M\rightarrow i\mu$, must be taken.

Similarly in other cases a proof of unitarity for Feynman propagators,
based on the decomposition (72), can be converted into a proof of
unitarity for Wheeler propagators, by using the corresponding
decomposition (78).

\section{Discussion}

We have shown that the Wheeler propagator has several interesting
properties. In the first place we have the fact that it implies only 
virtual propagation. The on-shell $\delta$-function, solution of the free
equation, is absent. No quantum of the field can be found in a free state.
The function is allways zero for space-like distances. The field
propagation takes place inside the light-cone. This is true for bradyons,
but is also true for fields that obey Klein-Gordon equations with the
wrong sign of the mass term and even for complex mass fields. The 
convolution of two Wheeler functions gives another Wheeler function.
In p-space, this convolution coincides for $P<0$, with the convolution
of the two on-shell $\delta$-functions. In spite of the fact that
each of the latter only contains free propagation, while each of the
former only contains virtual propagation. The usual vacuum state is
annihilated by the descending operator $a$, and gives rise to the Feynman
propagator. The Wheeler Green function is associated to the 
symmetric vacuum. This vacuum is not annihilated by $a$, but rather by the
``true�� energy'' operator, a symmetric combination of annihilation and                                                                     
creation operators. The space of states generated by a and $a^{+}$ has 
an indefinite metric. There are known methods to deal with this kind
of space. In particular we can define and handle all scalar products
by means of the ``holomorphic representation''\cite{tp14}. 
Due to the absence of asymptotic free waves, no Wheeler particle will 
appear in external legs of the Feynman diagrams. Only the propagator
will appear explicitly, associated to internal lines. So the theoretical 
tools to deal with matrix elements in spaces with indefinite metric,
will not, in actual facts be necessary for the evaluation of 
cross-sections. However, the descomposition of unity for spaces with
indefinite metric, is needed for the proof of another important point.
The inclusion of Wheeler fields and the correspondings Wheeler
propagators do not produce any violation of unitarity, when only normal
particles are found in external legs of Feynman diagrams.

To complete the theoretical framework for a rigorous mathematical 
analysis, it is perhaps convenient to notice that the propagators
we have defined, are continuous linear functionals on the space of
the entire analytic functions rapidly decreasing on the real axis.
They are known in general as ``Tempered Ultradistributions''
\cite{tp15,tp16,tp17,tp18}. The Fourier transformed space contains
the usual distributions and also admits exponentially increasing
functions (distributions of the exponential type)(see also 
ref.\cite{tp19}).

We must also answer the important question. What are the possible
uses of the Wheeler propagators?.

In the first place we would like to stress the fact that the
quanta of Wheeler fields can not appear as free particles. They can
only be detected as virtual mediators of interactions. It is in the
light of this observation that we must look for probable applications.

We will first take the case of a tachyon field. 
It is known that unitarity can not be achieved, provided we accept the 
implicit premise that only Feynman propagators are to be used, with the 
consequent presence of free tachyons. This
work can also be considered to be a proof of the incompatibility of
unitarity and Feynman propagator for tachyons. To this observation we 
add the fact that if the propagator is a Wheeler function, a tachyon
can not propagate freely. Consequently we are led to the acceptance 
of the complete spectrum: $\vec{p}^{\,2} < \mu^2$ and 
$\vec{p}^{\,2}\geq\mu^2$. With the remark that the real exponentials 
one gets for $\vec{p}^{\,2}<{\mu}^2$ are not eigenfunctions of the 
Hamiltonian (ref.\cite{tp8,tp21}). Furthermore, this procedure fits
naturally into the treatment for complex mass fields of section 5.
To this case it could be related the Higgs problems. The scalar field of
the standard model behaves as a tachyon field for low amplitudes. The
fact that the Higgs has not yet been observed suggest the posibility
that the corresponding propagator might be a Wheeler function 
\cite{tp22}. It is easy to see that this assumption does not spoil
any of the experimental confirmations of the model (On the contrary,
it adds the non observation of the free Higgs).

Another possible application appear in higher order equations. 
Those equations appear for example in some supersymmetric models for 
higher dimensional spaces \cite{tp6}. They
can be decompossed into Klein-Gordon factors with general mass
parmeters. The corresponding fields have Wheeler functions as 
propagators. It is interesting that there are models of higher order
equations, coupled to electromagnetism, which can be shown to be 
unitary, no matter how hig the order is \cite{tp23}. 

The acceptance of thachyons as Wheeler particles, might be of interest
for the bosonic string theory. With the symmetric vacuum we can show
that the Virasoro algebra turns out to be free of anomalies in
spaces of arbitrary number of dimensions \cite{tp24}. The excitations
of the string are Wheeler functions in this case.

\pagebreak

\end{document}